\documentclass[a4paper]{article}
\usepackage{authblk}  
\usepackage{cite}
\usepackage{url}
\usepackage{graphicx}
\usepackage{nameref,hyperref}

\usepackage{multirow,array}
\usepackage{color}  

\newenvironment{red-color}{\par\color{red}}{\par}

\DeclareGraphicsExtensions{.pdf,.png}

\begin{document}

\title{\Large {\bf ABL: An original active blacklist based on a modification of the SMTP}}

\author{\normalsize
Pablo M.\ Oliveira\textsuperscript{1}, 
Mateus B.\ Vieira\textsuperscript{2}, \\
Isaac C.\ Ferreira\textsuperscript{3},
Jo\~{a}o P.\ R.\ R.\ Leite\textsuperscript{1},
Edvard M.\ Oliveira\textsuperscript{1}, \\
Bruno T.\ Kuehne\textsuperscript{1},
Edmilson M.\ Moreira\textsuperscript{1*},
Ot\'{a}vio A.\ S.\ Carpinteiro\textsuperscript{1}}

\affil{\normalsize 
\textbf{1} Research Group on Systems and Computer Engineering, Federal University of Itajub\'{a}, Av.\ BPS 1303, Itajub\'{a}, MG, 37500--903, Brazil \\
\textbf{2} Flapper Tecnologia, R.\ Sergipe, 1440, 7\textsuperscript{o} andar, Belo Horizonte, MG, 30112--011, Brazil \\
\textbf{3} TRICOD Equipamentos Eletr\^{o}nicos Ind\'{u}stria e Com\'{e}rcio LTDA, R.\ Cel.\ Francisco Braz 185, Sala 21, Itajub\'{a}, MG, 37500--005, Brazil \\
\small \it \textbf{*} Corresponding author: Edmilson M.\ Moreira (e-mail: edmarmo@unifei.edu.br)
}

\date{August, 2022}
\maketitle
\thispagestyle{empty}

\section*{Abstract}

\noindent This paper presents a novel Active Blacklist (ABL) based on a modification of the Simple Mail Transfer Protocol (SMTP). ABL was implemented in the Mail Transfer Agent (MTA) Postfix of the e-mail server Zimbra and assessed exhaustively in a series of experiments. The modified server Zimbra showed computational performance and costs similar to those of the original server Zimbra when receiving legitimate e-mails. When receiving spam, however, it showed better computing performance and costs than the original Zimbra. Moreover, there was a considerable computational cost on the spammer's server when it sent spam e-mails. ABL was assessed at the Federal University of Itajub\'{a}, Brazil, during a period of sixty-one days. It was responsible for rejecting a percentage of 20.94\% of the spam e-mails received by the university during this period. After this period, it was deployed and remained in use, from July-2015 to July-2019, at the university. ABL is part of the new Open Machine-Learning-Based Anti-Spam (Open-MaLBAS). Both ABL and Open-MaLBAS are freely available on GitHub.

\vspace{3ex}
\Large\noindent\textbf{Keywords}
\vspace{1.5ex}

\normalsize\noindent Electronic mail (e-mail), unsolicited electronic mail (spam), simple mail transfer protocol (SMTP), internet, network security, open source software

\section{Introduction}
\label{sec:intro}

E-mail is the most popular communication service of the Internet. Unfortunately it is unduly exploited by spammers, individuals who send spam e-mails. Spam e-mails, or simply spam, are messages sent without previous consent of recipients \cite{geor14}. They usually convey publicity or malicious content.

The amount of spam has grown rapidly through the years. According to Statista Co., over 50\% of e-mail traffic circulating in the Internet consists in some kind of spam \cite{stat22}. Spam e-mails cause serious losses to institutions, as they overload servers, communications links and network appliances.

Any implementation of e-mail service should meet Simple Mail Transfer Protocol (SMTP) specifications \cite{klen15}. SMTP provides that any e-mail unduly sent to a non-existent address should be returned to the sender. This SMTP specification produces a behavior common to spammers. They remove from their distribution lists each e-mail address that is occasionally deactivated, in order not to bear the cost of returned e-mails.

This common behavior among spammers was clearly and significantly noted at the Federal University of Itajub\'{a}, Bra\-zil. A few years ago, the university deactivated several of its sub-domains. The e-mail servers of these sub-domains used to receive large volumes of spam. A few weeks later there was a need to reactivate the e-mail server of one of the sub-domains. This server remained active for two weeks. During this period it received no spam e-mail.

Slettnes \cite{slet22} says it is advantageous to filter e-mail during an incoming SMTP transaction, rather than following the conventional approach of leaving this task to an anti-spam system. He lists some advantages, such as: (a) the possibility to stop the delivery of most spam early in the SMTP transaction, before the actual message data has been received, saving both network bandwidth and CPU processing; (b) the possibility to deploy some spam filtering techniques that are not possible later (e.g., SMTP transaction delays, blacklisting, and greylisting); and (c) the possibility to notify the sender in case of a delivery failure (e.g., owing to an invalid recipient address) without directly generating collateral spam.

Owing to these advantages, current MTAs are usually capable to refuse spam during the SMTP transaction if properly configured. However, it is worth noticing that the refusal in receiving spam during the SMTP transaction does not prevent the spammer either from finding new methods to avoid the anti-spam configuration of MTAs or from going on sending spam to the same recipient in the future.

This paper presents an original Active Blacklist (ABL) based on a modification of the Simple Mail Transfer Protocol (SMTP). It differs from usual passive blacklists in that it produces three advantageous consequences. The first consists in promptly rejecting, during SMTP negotiation, the undesired (i.e., spam) e-mails thus defined by each recipient (i.e., e-mail user) of an institution, avoiding a waste of the computational (e.g., CPU, memory, storage) and network (e.g., bandwith, network appliances) resources of the institution. The second consists in the return of the undesired e-mails to the sender (i.e., spammer), penalizing him/her as his/her server will use more computational and network resources to handle the rejected e-mails. In other words, ABL rejects the spam e-mail {\it before\/} receiving it, during SMTP negotiation, wasting the spammer's resources, whereas any usual passive blacklist rejects the spam e-mail {\it after\/} receiving it completely, wasting the resources of the institution. The third consequence consists in the fact that owing to the cost of the refusal, the sender (i.e., spammer) removes the recipient from his/her distribution lists.

ABL targets specifically spam messages with fixed sender addresses. It is worth noticing, however, that many spam e-mails originate from mailing lists of e-commerce firms, which have fixed sender addresses. On making a purchase from these firms, for instance, the customer has his/her e-mail automatically recorded without his/her consent in their mailing lists, and often also in those of their partners. As observed by customers, many e-commerce firms do not remove their e-mail addresses from their lists even when requested by them. Reducing this undesired traffic in itself justifies implementing ABL.

ABL is part of the new Open Machine-Learning-Based Anti-Spam (Open-MaLBAS) \cite{malb21}. Both ABL \cite{abl22} and Open-MaLBAS \cite{malb22} are freely available on GitHub.

This paper is divided into nine sections. The second section presents the novel ABL. The third and fourth sections describe respectively the hardware and software resources employed or developed to carry out the experiments. The fifth section presents the experiments. The results obtained are assessed in the sixth section. The seventh section describes and evaluates the deployment of ABL in the production environment of the Federal University of Itajub\'{a}. The eighth section contrasts ABL with current anti-spam mechanisms. Finally, the ninth section puts forth final considerations and shows directions for future research.

\section{Active blacklist (ABL)}
\label{sec:abl}

\begin{figure}[htb]
\centering
\includegraphics[width=45mm]{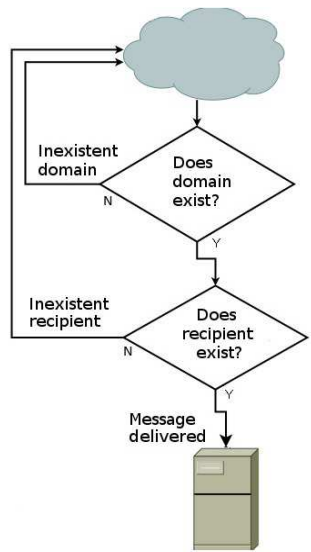}
\caption{\bf E-mail receiving process defined by SMTP}
\label{fig:smtp}
\end{figure}

During the e-mail receiving process as defined by SMTP, the recipient's domain and address are checked. Figure~\ref{fig:smtp} describes the receiving process.

\begin{figure}[htb]
\centering
\includegraphics[width=45mm]{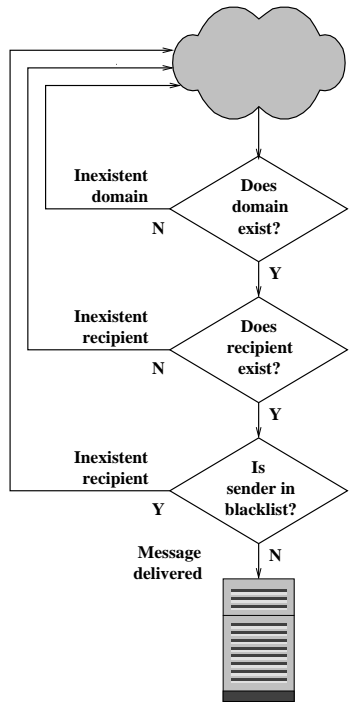}
\caption{\bf E-mail receiving process defined by ABL}
\label{fig:abl}
\end{figure}

ABL modifies the e-mail receiving process, adding one further verification. This verification consists in consulting a list of undesired e-mails (i.e., blacklist) thus defined by the recipient (i.e., e-mail user), in order to check the existence of the e-mail address of the sender. Should it exist, the e-mail will be refused, returning the message ``Undelivered Mail Returned to Sender'' to the sender. Figure~\ref{fig:abl} describes the e-mail receiving process defined by ABL.

The module {\it smtpd\/} of the Mail Transfer Agent (MTA) Postfix 2.7.5 \cite{post22} was modified to contain the ABL. Postfix 2.7.5 is included in the e-mail server Zimbra 8.0 \cite{zimb22}. The modification in the module {\it smtpd\/} consisted in creating the function {\it check\_spam\/} in the file {\it smtpd\_check.c\/}. The function {\it check\_spam\/} checks the existence of the sender's address in the recipient's blacklist. It is called through the Postfix function {\it reject\_unknown\_address\/}, which checks whether the domain and recipient exist. In all, 30 lines were included in the Postfix original source code.

The blacklist is implemented as a table named {\it BlackList}, in the Zimbra Database Management System (DBMS) MySQL 5.1 \cite{mysq22}. MySQL uses the InnoDB engine for performance reasons.

\section{Hardware resources}
\label{sec:hard}

\begin{figure}[htb]
\centering
\includegraphics[width=45mm]{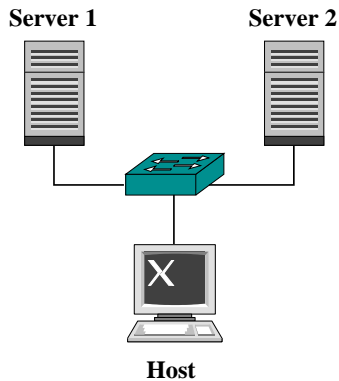}
\caption{\bf Hardware architecture employed in the experiments}
\label{fig:hard}
\end{figure}

The hardware architecture employed in the experiments is described in Figure~\ref{fig:hard}. It consists of two identical servers DELL T110, with 2GHz quad-core Intel Xeon processor and 32GB RAM memory, and a host with 3GHz Pentium 4 processor and 512MB RAM memory. The servers and the host use the Linux Ubuntu Server 10.04 operating system. E-mails are simultaneously sent from the host to both servers.

The host and the servers run the e-mail server Zimbra 8.0, which includes the Postfix 2.7.5 as MTA. The Zimbra anti-spam (Edge MTA) and antivirus filters were deactivated. The Postfix module {\it smtpd\/} which runs in server 2 was modified to contain ABL. Server 1 and the host run the Postfix original code.

\section{Software resources}
\label{sec:soft}

\subsection{Ancillary tools}

Several ancillary tools were created to collect, process and format data. Some of these tools are described below.

The tool {\it monitor\/} monitors the uptime, CPU usage, RAM memory usage, transmission (Tx) and reception (Rx) of bytes in the network interface in the three computers. It also monitors the size of queues {\it incoming}, {\it maildrop}, {\it active}, {\it deferred}, {\it hold\/} and {\it corrupt\/} in both servers Zimbra. Data monitored in each computer are saved in a file {\it logmail}. The tool is run every two seconds by means of a daemon.

The tool {\it send-email\/} is run in the host. It performs two tasks. The first consists in initializing the file {\it logmail\/} in each computer. The second task consists in sending e-mails from the host to both servers Zimbra, in a synchronized form. The e-mails of random sizes are created by the host and sent to its MTA Postfix by means of SMTP command sequences inside a telnet connection in the port 25. Two {\it logfiles\/} are created to record the success or failure in sending each e-mail to the servers. The lines of a configuration file, known as {\it catalog\/}, define the number of e-mails for sending, the number of recipients and of senders.

The tool {\it verify-delay\/} monitors the time spent by each server to deliver the e-mails to recipients. It uses data stored in file {\it mailbox.log}, created by the Zimbra. This file records events occurred in the mailboxes. E-mail delivery time is the only event of interest.

The tool {\it summary\/} summarizes the results of experiments. It takes data stored in the files {\it logmail\/} of the three computers and calculates the average rate of CPU usage, RAM memory usage, transmission (Tx) and reception (Rx) of bytes in the network interface. It also calculates time spent to deliver all the e-mails to mailboxes and the maximum number of e-mails found in the queue {\it incoming\/} of each server.

\subsection{Blacklist interface}
\label{sec:interface}

An interface was devised in order to let recipients (i.e., e-mail users) record, list and delete, in a simple form, e-mail addresses deemed to be spam. The interface was developed by means of a plug-in for the Mozilla Thunderbird e-mail client \cite{thun22}. The Thunderbird plug-in \cite{plug22} is freely available on GitHub.

\begin{figure}[htb]
\centering
\includegraphics[width=100mm]{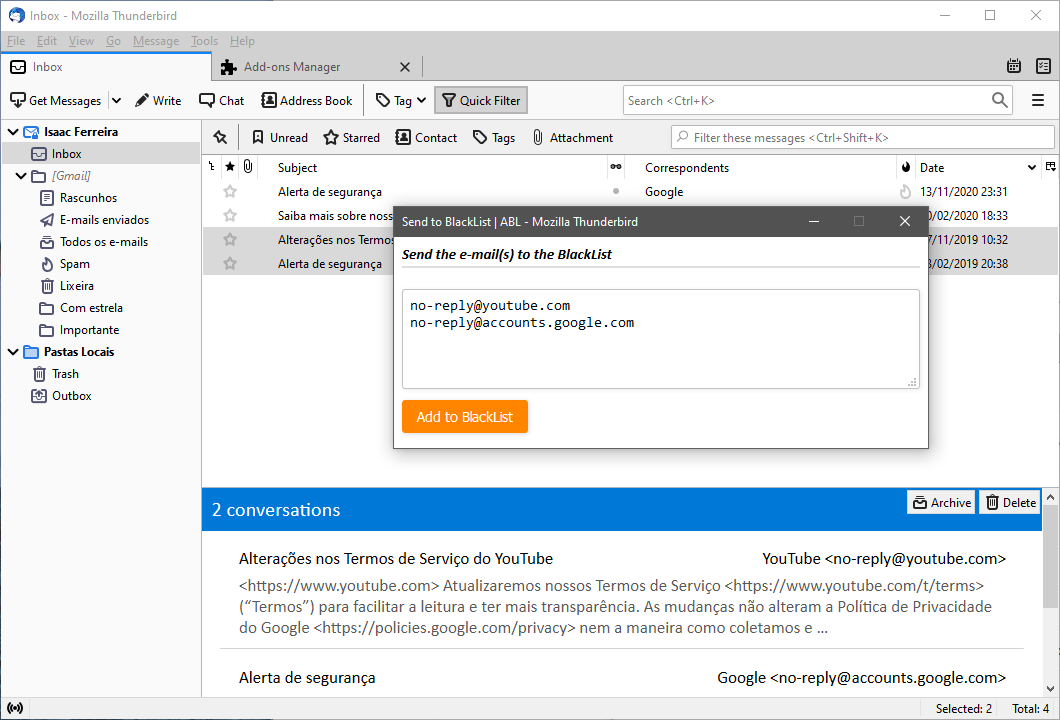}
\caption{\bf Plug-in interface --- list of e-mails for blocking}
\label{fig:iface-1}
\end{figure}

\begin{figure}[htb]
\centering
\includegraphics[width=100mm]{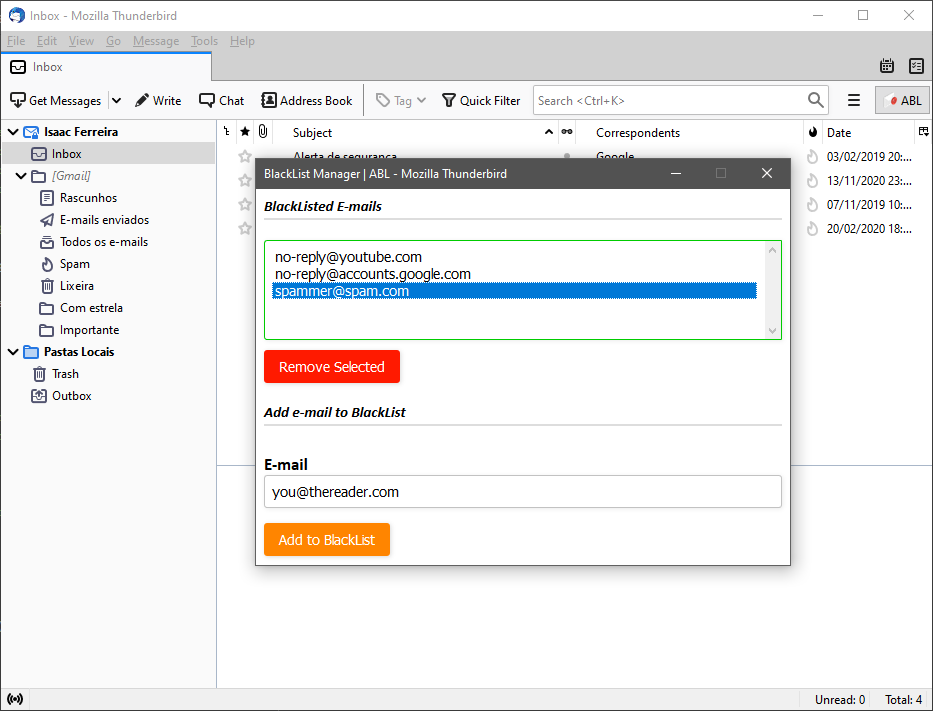}
\caption{\bf Plug-in interface --- list of blocked e-mails}
\label{fig:iface-2}
\end{figure}

The plug-in, written in JavaScript, communicates with the server Zimbra through a web service. The web service, written in PHP, contains the credentials to communicate with the Zimbra DBMS. Hence, the plug-in may be freely distributed, as it contains no confidential information. Figures~\ref{fig:iface-1} and \ref{fig:iface-2} show the plug-in interface.

\section{Experiments}

The experiments are intended to assess the performance of the modified server Zimbra (server 2) in relation to that of the original server Zimbra (server 1). The server performances were measured by means of six metrics --- average rate of CPU usage, average rate of RAM memory usage, average rate of bytes transmitted (Tx) in the network interface, average rate of bytes received (Rx) in the network interface, maximum number of e-mails found in the queue {\it incoming\/} and time spent to deliver all the e-mails to mailboxes.

Server 1 runs the Postfix original module {\it smtpd\/} and server 2 runs the modified one that contains the ABL. The host sends simultaneously 100, 1,000, 10,000 and 100,000 e-mails to each server. These e-mails sent are either ham\footnote{Legitimate e-mails are also referred to as hams.} only, spam only, or a combination of ham and spam. E-mails are sent either to one or to 100 recipients. Each recipient has either 30, 78 or 100 addresses registered as spam in the MySQL {\it BlackList\/} table.

As server 2 does not deliver spam e-mails to recipients, the time spent in e-mail delivery is not measured in experiments 5 to 10, which involve sending and delivering spam e-mails. In experiments 9 and 10, 78\% of e-mails sent by the host are spam. This percentage is equal to the average percentage of spam the university's server receives. Table~\ref{tab:exp-desc} describes the parameters employed in each of the ten experiments.

\begin{table}[htb]
\begin{center}
\setlength{\extrarowheight}{2pt}
\caption{\bf Parameters employed in experiments --- R(s): recipient(s); SAs: spam addresses}
\label{tab:exp-desc}
\vspace{1ex}
\begin{tabular}{|c||c|c|c|c|c|} \hline
Experi-\ & No.     & E-mail & No.   & No.  & Total \\
ment\    & e-mails & type   & Rs & SAs/R & SAs \\ \hline \hline
 1 & 100/1K/10K/100K & ham & 1 &  30 & 30  \\ \hline
 2 & 100/1K/10K/100K & ham & 1 & 100 & 100 \\ \hline
 3 & 100/1K/10K/100K & ham & 100 & 30 & 3000 \\ \hline
 4 & 100/1K/10K/100K & ham & 100 & 100 & 10000 \\ \hline
 5 & 100/1K/10K/100K & spam & 1 & 30 & 30 \\ \hline
 6 & 100/1K/10K/100K & spam & 1 & 100 & 100 \\ \hline
 7 & 100/1K/10K/100K & spam & 100 & 30 & 3000 \\ \hline
 8 & 100/1K/10K/100K & spam & 100 & 100 & 10000 \\ \hline
 9 & 100/1K/10K/100K & ham/spam & 1 & 78 & 78 \\ \hline
10 & 100/1K/10K/100K & ham/spam & 100 & 78 & 7800 \\ \hline
\end{tabular}
\end{center}
\end{table}


\section{Assessment of the experiments}

\subsection{Assessment of experiments 1 to 4}

Experiments 1 to 4 involve sending ham e-mails only to the servers. Server 2 checks the MySQL table {\it BlackList\/} before delivering each ham e-mail to its respective mailbox. Therefore, it is expected that the results obtained by server 2 be slightly below those obtained by server 1.

Indeed, in all six metrics server 2 achieved performance slightly below that of server 1. The rate of CPU usage of server 2 was on average 6.66\% above that of server 1. The rates of RAM memory usage, of bytes transmitted, and of bytes received by server 2 were on average, respectively 1.22\%, 2.22\% and 0.52\% above those of server 1. The maximum number of e-mails found in the queue {\it incoming\/} of server 2 and the time it spent to deliver all the e-mails to the mailboxes grew on average 0.61\% and 0.01\% respectively.

The results indicate that checking the {\it BlackList\/} table before accepting ham e-mails adds insignificant computational costs to the modified server Zimbra. The computational costs of one server in relation to those of the other, measured by means of each of the six metrics, are all described in Table~\ref{tab:comp1-4}.

\begin{table}[htb]
\begin{center}
\setlength{\extrarowheight}{2pt}
\caption{\bf Results of one server in relation to those of the other --- experiments 1--4}
\label{tab:comp1-4}
\vspace{1ex}
\begin{tabular}{|c||c|c|c|c|c|c|c|} \hline
\multicolumn{7}{|c|}{Experiment 1} \\ \cline{1-7}
\multicolumn{1}{|c||}{No.\ e-} &
\multicolumn{1}{c|}{CPU} &
\multicolumn{1}{c|}{RAM} &
\multicolumn{1}{c|}{Tx} &
\multicolumn{1}{c|}{Rx} &
\multicolumn{1}{c|}{Queue} &
\multicolumn{1}{c|}{Deliv.} \\
mails & {\scriptsize (S2/S1)} & {\scriptsize (S2/S1)} & {\scriptsize (S2/S1)} & {\scriptsize (S2/S1)} & {\scriptsize (S2/S1)} & {\scriptsize (S2/S1)} \\ \hline \hline
100 & 1.0000 & 1.0045 & 1.0698 & 1.0026 & 1.0000 & 1.0000 \\ \hline
1000 & 1.0833 & 1.0024 & 1.0260 & 1.0009 & 1.0000 & 1.0000 \\ \hline
10000 & 1.0833 & 1.0101 & 1.0213 & 1.0003 & 1.0306 & 1.0001 \\ \hline
100000 & 1.0175 & 1.0158 & 1.0037 & 1.0010 & 1.0040 & 1.0001 \\ \hline \hline \hline
\multicolumn{7}{|c|}{Experiment 2} \\ \cline{1-7}
\multicolumn{1}{|c||}{No.\ e-} &
\multicolumn{1}{c|}{CPU} &
\multicolumn{1}{c|}{RAM} &
\multicolumn{1}{c|}{Tx} &
\multicolumn{1}{c|}{Rx} &
\multicolumn{1}{c|}{Queue} &
\multicolumn{1}{c|}{Deliv.} \\
mails & {\scriptsize (S2/S1)} & {\scriptsize (S2/S1)} & {\scriptsize (S2/S1)} & {\scriptsize (S2/S1)} & {\scriptsize (S2/S1)} & {\scriptsize (S2/S1)} \\ \hline \hline
100 & 1.0909 & 1.0045 & 1.0227 & 1.0026 & 1.0000 & 1.0000 \\ \hline
1000 & 1.0769 & 1.0023 & 1.0256 & 1.0009 & 1.0000 & 1.0000 \\ \hline
10000 & 1.1304 & 1.0100 & 1.0210 & 1.0003 & 1.0097 & 1.0001 \\ \hline
100000 & 1.0172 & 1.0158 & 1.0037 & 1.0010 & 1.0040 & 1.0001 \\ \hline \hline \hline
\multicolumn{7}{|c|}{Experiment 3} \\ \cline{1-7}
\multicolumn{1}{|c||}{No.\ e-} &
\multicolumn{1}{c|}{CPU} &
\multicolumn{1}{c|}{RAM} &
\multicolumn{1}{c|}{Tx} &
\multicolumn{1}{c|}{Rx} &
\multicolumn{1}{c|}{Queue} &
\multicolumn{1}{c|}{Deliv.} \\
mails & {\scriptsize (S2/S1)} & {\scriptsize (S2/S1)} & {\scriptsize (S2/S1)} & {\scriptsize (S2/S1)} & {\scriptsize (S2/S1)} & {\scriptsize (S2/S1)} \\ \hline \hline
100 & 1.0000 & 1.0125 & 1.0682 & 1.0106 & 1.0000 & 1.0000 \\ \hline
1000 & 1.0833 & 1.0103 & 1.0318 & 1.0089 & 1.0000 & 1.0000 \\ \hline
10000 & 1.0800 & 1.0182 & 1.0215 & 1.0082 & 1.0194 & 1.0001 \\ \hline
100000 & 1.0172 & 1.0239 & 1.0037 & 1.0090 & 1.0119 & 1.0001 \\ \hline \hline \hline
\multicolumn{7}{|c|}{Experiment 4} \\ \cline{1-7}
\multicolumn{1}{|c||}{No.\ e-} &
\multicolumn{1}{c|}{CPU} &
\multicolumn{1}{c|}{RAM} &
\multicolumn{1}{c|}{Tx} &
\multicolumn{1}{c|}{Rx} &
\multicolumn{1}{c|}{Queue} &
\multicolumn{1}{c|}{Deliv.} \\
mails & {\scriptsize (S2/S1)} & {\scriptsize (S2/S1)} & {\scriptsize (S2/S1)} & {\scriptsize (S2/S1)} & {\scriptsize (S2/S1)} & {\scriptsize (S2/S1)} \\ \hline \hline
100 & 1.0000 & 1.0125 & 0.9792 & 1.0106 & 1.0000 & 1.0000 \\ \hline
1000 & 1.1667 & 1.0103 & 1.0318 & 1.0089 & 1.0000 & 1.0000 \\ \hline
10000 & 1.1667 & 1.0182 & 1.0215 & 1.0082 & 1.0098 & 1.0001 \\ \hline
100000 & 1.0517 & 1.0239 & 1.0037 & 1.0090 & 1.0079 & 1.0001 \\ \hline \hline \hline
\multicolumn{7}{|c|}{{\bf M E A N}} \\ \cline{1-7}
\multicolumn{1}{|c||}{} &
\multicolumn{1}{c|}{CPU} &
\multicolumn{1}{c|}{RAM} &
\multicolumn{1}{c|}{Tx} &
\multicolumn{1}{c|}{Rx} &
\multicolumn{1}{c|}{Queue} &
\multicolumn{1}{c|}{Deliv.} \\ \hline \hline
{\bf Mean} & {\it 1.0666} & {\it 1.0122} & {\it 1.0222} & {\it 1.0052} & {\it 1.0061} & {\it 1.0001} \\ \hline
\end{tabular}
\end{center}
\end{table}


\subsection{Assessment of experiments 5 to 8}

Experiments 5 to 8 involve sending spam e-mails only to the servers. Server 2 refuses spam e-mails, returning the message ``Undelivered Mail Returned to Sender'' to the sender. Hence, it is expected that the results obtained by server 2 be below those obtained by server 1 only in terms of the average rate of bytes transmitted (Tx) in the network interface. On the other hand, the results obtained by server 2 are expected to be above or at least similar to those obtained by server 1 in all other metrics.

Indeed, server 2 transmits to the host on average 720.57 times more bytes than server 1. On the other hand, server 1 receives from the host on average 641.62 times more bytes than server 2. Moreover, server 1 uses on average 2.36 times more CPU than server 2, and the maximum size achieved by its queue {\it incoming\/} is on average 231.88 times greater than the maximum size achieved by that of server 2. Average rates of RAM memory usage in both servers were similar.

The results indicate that on rejecting spam e-mails, server 2 reduces significantly the use of its CPU as well as the maximum number of e-mails in its queue {\it incoming}. Transmission (Tx) of bytes through its network interface increases owing to the immediate return of spam e-mails to senders. Nonetheless, this increase is offset by the reduction in the number of bytes received (Rx) in its network interface. The computational costs of one server in relation to those of the other, measured by means of each of the five metrics, are all described in Table~\ref{tab:comp5-8}.

\begin{table}[htb]
\begin{center}
\setlength{\extrarowheight}{2pt}
\caption{\bf Results of one server in relation to those of the other --- experiments 5--8}
\label{tab:comp5-8}
\vspace{1ex}
\begin{tabular}{|c||c|c|c|c|c|c|c|} \hline
\multicolumn{6}{|c|}{Experiment 5} \\ \cline{1-6}
\multicolumn{1}{|c||}{No.\ e-} &
\multicolumn{1}{c|}{CPU} &
\multicolumn{1}{c|}{RAM} &
\multicolumn{1}{c|}{Tx} &
\multicolumn{1}{c|}{Rx} &
\multicolumn{1}{c|}{Queue} \\
mails & {\scriptsize (S1/S2)} & {\scriptsize (S2/S1)} & {\scriptsize (S2/S1)} & {\scriptsize (S1/S2)} & {\scriptsize (S1/S2)} \\ \hline \hline
100 & 1.0000 & 1.0248 & 296.10 & 124.56 & 1.00 \\ \hline
1000 & 0.9231 & 1.0055 & 794.15 & 760.47 & 1.00 \\ \hline
10000 & 2.4545 & 1.0218 & 926.13 & 836.25 & 99.00 \\ \hline
100000 & 5.3636 & 0.9708 & 864.04 & 845.54 & 813.00 \\ \hline \hline \hline
\multicolumn{6}{|c|}{Experiment 6} \\ \cline{1-6}
\multicolumn{1}{|c||}{No.\ e-} &
\multicolumn{1}{c|}{CPU} &
\multicolumn{1}{c|}{RAM} &
\multicolumn{1}{c|}{Tx} &
\multicolumn{1}{c|}{Rx} &
\multicolumn{1}{c|}{Queue} \\
mails & {\scriptsize (S1/S2)} & {\scriptsize (S2/S1)} & {\scriptsize (S2/S1)} & {\scriptsize (S1/S2)} & {\scriptsize (S1/S2)} \\ \hline \hline
100 & 1.0909 & 1.0248 & 299.05 & 124.52 & 1.00 \\ \hline
1000 & 1.0000 & 1.0054 & 796.70 & 758.77 & 1.00 \\ \hline
10000 & 2.0769 & 1.0218 & 926.27 & 836.23 & 100.00 \\ \hline
100000 & 4.6154 & 0.9708 & 864.03 & 845.55 & 821.00 \\ \hline \hline \hline
\multicolumn{6}{|c|}{Experiment 7} \\ \cline{1-6}
\multicolumn{1}{|c||}{No.\ e-} &
\multicolumn{1}{c|}{CPU} &
\multicolumn{1}{c|}{RAM} &
\multicolumn{1}{c|}{Tx} &
\multicolumn{1}{c|}{Rx} &
\multicolumn{1}{c|}{Queue} \\
mails & {\scriptsize (S1/S2)} & {\scriptsize (S2/S1)} & {\scriptsize (S2/S1)} & {\scriptsize (S1/S2)} & {\scriptsize (S1/S2)} \\ \hline \hline
100 & 1.0000 & 1.0248 & 294.48 & 124.46 & 1.00 \\ \hline
1000 & 0.9231 & 1.0054 & 793.94 & 761.66 & 1.00 \\ \hline
10000 & 2.3333 & 1.0218 & 926.41 & 836.22 & 101.00 \\ \hline
100000 & 5.0000 & 0.9709 & 864.02 & 845.56 & 829.00 \\ \hline \hline \hline
\multicolumn{6}{|c|}{Experiment 8} \\ \cline{1-6}
\multicolumn{1}{|c||}{No.\ e-} &
\multicolumn{1}{c|}{CPU} &
\multicolumn{1}{c|}{RAM} &
\multicolumn{1}{c|}{Tx} &
\multicolumn{1}{c|}{Rx} &
\multicolumn{1}{c|}{Queue} \\
mails & {\scriptsize (S1/S2)} & {\scriptsize (S2/S1)} & {\scriptsize (S2/S1)} & {\scriptsize (S1/S2)} & {\scriptsize (S1/S2)} \\ \hline \hline
100 & 1.0000 & 1.0248 & 297.35 & 124.41 & 1.00 \\ \hline
1000 & 0.9231 & 1.0056 & 796.45 & 759.98 & 1.00 \\ \hline
10000 & 2.5455 & 1.0219 & 925.91 & 836.20 & 102.00 \\ \hline
100000 & 5.5455 & 0.9708 & 864.06 & 845.52 & 837.00 \\ \hline \hline \hline
\multicolumn{6}{|c|}{{\bf M E A N}} \\ \cline{1-6}
\multicolumn{1}{|c||}{} &
\multicolumn{1}{c|}{CPU} &
\multicolumn{1}{c|}{RAM} &
\multicolumn{1}{c|}{Tx} &
\multicolumn{1}{c|}{Rx} &
\multicolumn{1}{c|}{Queue} \\ \hline \hline
{\bf Mean} & {\it 2.3622} & {\it 1.0057} & {\it 720.57} & {\it 641.62} & {\it 231.88} \\ \hline
\end{tabular}
\end{center}
\end{table}

\clearpage

\subsection{Assessment of experiments 9 to 10}

Experiments 9 to 10 involve sending 22\% in ham e-mails and 78\% in spam e-mails to the servers. This percentage is equal to the average percentage of spam the university's server receives. It is thus expected that the results obtained by the servers be found between the values of the results of experiments 1--4 and those of the results of experiments 5--8.

Indeed, in four of the metrics --- average rate of CPU usage, average rates of bytes transmitted (Tx) and received (Rx) in the network interface, and maximum number of e-mails found in the queue {\it incoming\/} --- the results were found between those of experiments 1--4 and those of experiments 5--8. Average rates of RAM memory usage in both servers were similar to those of the other experiments. The computational costs of one server in relation to those of the other, measured by means of each of the five metrics, are all described in Table~\ref{tab:comp9-10}.

\begin{table}[htb]
\begin{center}
\setlength{\extrarowheight}{2pt}
\caption{\bf Results of one server in relation to those of the other --- experiments 9--10}
\label{tab:comp9-10}
\vspace{1ex}
\begin{tabular}{|c||c|c|c|c|c|c|c|} \hline
\multicolumn{6}{|c|}{Experiment 9} \\ \cline{1-6}
\multicolumn{1}{|c||}{No.\ e-} &
\multicolumn{1}{c|}{CPU} &
\multicolumn{1}{c|}{RAM} &
\multicolumn{1}{c|}{Tx} &
\multicolumn{1}{c|}{Rx} &
\multicolumn{1}{c|}{Queue} \\
mails & {\scriptsize (S1/S2)} & {\scriptsize (S1/S2)} & {\scriptsize (S2/S1)} & {\scriptsize (S1/S2)} & {\scriptsize (S1/S2)} \\ \hline \hline
100 & 1.0000 & 0.9647 & 229.06 & 2.56 & 1.00 \\ \hline
1000 & 0.9167 & 0.9669 & 618.12 & 3.18 & 1.00 \\ \hline
10000 & 1.3333 & 0.9645 & 722.14 & 3.19 & 1.36 \\ \hline
100000 & 1.3590 & 1.0692 & 673.94 & 3.19 & 1.37 \\ \hline \hline \hline
\multicolumn{6}{|c|}{Experiment 10} \\ \cline{1-6}
\multicolumn{1}{|c||}{No.\ e-} &
\multicolumn{1}{c|}{CPU} &
\multicolumn{1}{c|}{RAM} &
\multicolumn{1}{c|}{Tx} &
\multicolumn{1}{c|}{Rx} &
\multicolumn{1}{c|}{Queue} \\
mails & {\scriptsize (S1/S2)} & {\scriptsize (S1/S2)} & {\scriptsize (S2/S1)} & {\scriptsize (S1/S2)} & {\scriptsize (S1/S2)} \\ \hline \hline
100 & 1.0000 & 1.0295 & 221.54 & 3.24 & 1.00 \\ \hline
1000 & 0.9167 & 1.0321 & 591.67 & 3.32 & 1.00 \\ \hline
10000 & 1.3684 & 1.0297 & 691.91 & 3.32 & 1.37 \\ \hline
100000 & 1.3333 & 1.1412 & 645.56 & 3.32 & 1.37 \\ \hline \hline \hline
\multicolumn{6}{|c|}{{\bf M E A N}} \\ \cline{1-6}
\multicolumn{1}{|c||}{} &
\multicolumn{1}{c|}{CPU} &
\multicolumn{1}{c|}{RAM} &
\multicolumn{1}{c|}{Tx} &
\multicolumn{1}{c|}{Rx} &
\multicolumn{1}{c|}{Queue} \\ \hline \hline
{\bf Mean} & {\it 1.1534} & {\it 1.0247} & {\it 549.24} & {\it 3.16} & {\it 1.18} \\ \hline
\end{tabular}
\end{center}
\end{table}


\subsection{Performance of the host}

In all the experiments, the rates of CPU usage, of RAM memory usage, and of transmission (Tx) and reception (Rx) of bytes in the host's network interface were measured. Table~\ref{tab:exp-host} shows the average rates obtained by the host in the ten experiments, classed by the type of e-mail sent as well as by the number of e-mails sent.

\begin{table}[htb]
\begin{center}
\setlength{\extrarowheight}{2pt}
\caption{\bf Results of the host in the ten experiments}
\label{tab:exp-host}
\vspace{1ex}
\begin{tabular}{|c||c|c|c|c|c|} \hline
\multicolumn{5}{|c|}{Experiments 1--4 --- ham only} \\ \cline{1-5}
\multicolumn{1}{|c||}{No.\ e-} &
\multicolumn{1}{c|}{CPU} &
\multicolumn{1}{c|}{RAM} &
\multicolumn{1}{c|}{Tx} &
\multicolumn{1}{c|}{Rx} \\
mails & (\%) & (Kbyte) & (byte) & (byte) \\ \hline \hline
100 & 10 & 2624 & 27285 & 99 \\ \hline
1000 & 11 & 2673 & 277514 & 355 \\ \hline
10000 & 12 & 2740 & 3035048 & 3349 \\ \hline
100000 & 13 & 3271 & 28108821 & 35492 \\ \hline
{\bf Mean} & {\it 11.50} & {\it 2827.00} & {\it 7862167.00} & {\it 9823.75} \\ \hline \hline \hline
\multicolumn{5}{|c|}{Experiments 5--8 --- spam only} \\ \cline{1-5}
\multicolumn{1}{|c||}{No.\ e-} &
\multicolumn{1}{c|}{CPU} &
\multicolumn{1}{c|}{RAM} &
\multicolumn{1}{c|}{Tx} &
\multicolumn{1}{c|}{Rx} \\
mails & (\%) & (Kbyte) & (byte) & (byte) \\ \hline \hline
100 & 19 & 5211 & 13126 & 12491 \\ \hline
1000 & 21 & 5308 & 135165 & 126863 \\ \hline
10000 & 23 & 5442 & 1456066 & 1421225 \\ \hline
100000 & 26 & 6496 & 14126675 & 13971929 \\ \hline
{\bf Mean} & {\it 22.25} & {\it 5614.25} & {\it 3932758.00} & {\it 3883127.00} \\ \hline \hline \hline
\multicolumn{5}{|c|}{Experiments 9--10 --- ham/spam} \\ \cline{1-5}
\multicolumn{1}{|c||}{No.\ e-} &
\multicolumn{1}{c|}{CPU} &
\multicolumn{1}{c|}{RAM} &
\multicolumn{1}{c|}{Tx} &
\multicolumn{1}{c|}{Rx} \\
mails & (\%) & (Kbyte) & (byte) & (byte) \\ \hline \hline
100 & 18 & 4904 & 16343 & 8957 \\ \hline
1000 & 20 & 4996 & 167754 & 90910 \\ \hline
10000 & 22 & 5122 & 1807034 & 1018391 \\ \hline
100000 & 24 & 6114 & 17531646 & 10011938 \\ \hline
{\bf Mean} & {\it 21.00} & {\it 5284.00} & {\it 4880694.00} & {\it 2782549.00} \\ \hline
\end{tabular}
\end{center}
\end{table}


It is possible to see in the results described in Table~\ref{tab:exp-host} that there is a considerable computational cost to the host when it sends spam e-mails. When sending spam e-mails only, its average rates of CPU usage, of RAM memory usage, and of bytes received (Rx) through its network interface increase in 93.48\%, 98.59\% and in almost 40,000\% respectively, in relation to the average rates obtained when only ham e-mails are sent. Nonetheless, when sending spam e-mails only, the average rate of bytes transmitted (Tx) through its network interface decreases in 50\% in relation to the average rate obtained when only ham e-mails are sent.

\section{Deployment}

The ABL was deployed at the Federal University of Itajub\'{a}, Brazil. It remained in use from July-2015 to July-2019. From August-2019 on, the university e-mail service has been providing through the G-Suite platform of Google.

During the operational period of ABL, the university made use of both the e-mail server Zimbra 8.0 and the anti-spam filter CanIt-PRO 9.2.4 \cite{cani22}. As CanIt-PRO includes the Postfix 2.7.5, the university used the CanIt-PRO Postfix as its MTA. Thus, e-mails classified as legitimate by CanIt-PRO were sent directly to the e-mail server Zimbra in order to be delivered to the mailboxes of the recipients (i.e., e-mail users).

Two modifications were made in the CanIt-PRO. Firstly, its Postfix module {\it smtpd\/} was modified to contain the ABL. Secondly, it was installed in it a DBMS MySQL 5.1 with InnoDB engine.

The ABL also included a counter in its source code in order to count how many times it refused promptly (i.e., during SMTP negotiation) spam e-mails. The value of this counter was saved daily in a file and was reset immediately after.

CanIt-PRO sends daily an e-mail to each recipient, listing the recipient's e-mails which it put in quarantine. The recipient is thus able to mark these e-mails either as legitimate or as spam. If the recipient marks an e-mail as legitimate, it leaves the quarantine and is sent to the recipient. Otherwise, the e-mail is deleted from quarantine.

The deployment of the ABL made use of this CanIt-PRO procedure. The e-mails in quarantine with fixed sender addresses were used to populate the {\it BlackList\/} table of CanIt-PRO DBMS. Nonetheless, when a recipient marked an e-mail in quarantine as legitimate, its sender address was also removed from the DBMS {\it BlackList\/} table.

It is necessary to mention that the recipient (i.e., the e-mail user) is the {\it only\/} one responsible for creating, removing, and modifying his/her blacklist in the ABL. It is only him/her who decides which e-mails s/he wants to refuse. Therefore, the use of the CanIt-PRO procedure was provisional. It was provisionally used only to populate quickly the {\it BlackList\/} table of CanIt-PRO DBMS, enabling the faster production and report of the results.

The {\it BlackList\/} table of CanIt-PRO DBMS was then populated with 10,561 different recipients corresponding to the 10,561 staff members and students of the university. The table contained 54,917 sender addresses of spam, which give a rate of 5.2 sender addresses per recipient.

\begin{figure}[htb]
\centering
\includegraphics[width=120mm]{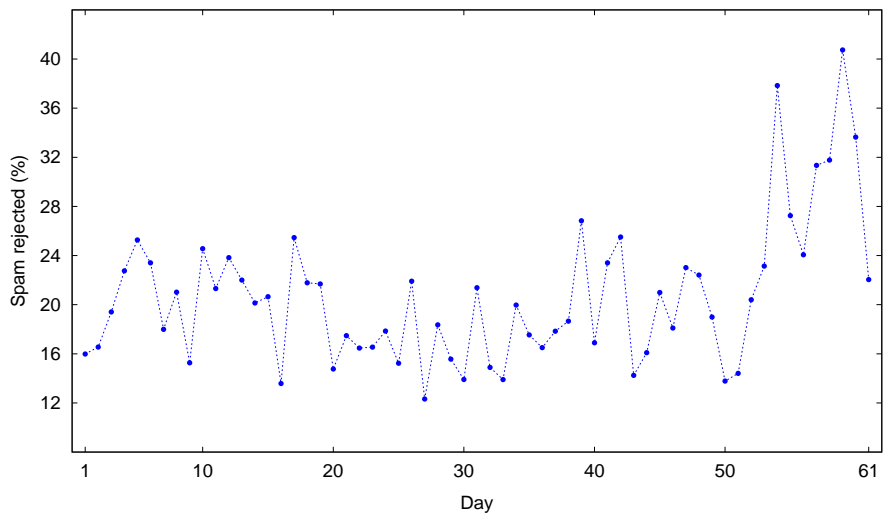}
\caption{\bf Daily percentage of spam e-mails rejected promptly (i.e., during SMTP negotiation) by ABL}
\label{fig:deploy}
\end{figure}

ABL was monitored during a period of sixty-one days. After that, ABL ran without monitoring. The results are very good, as expected. During the period, the university received a total of 2,402,429 e-mails. ABL refused promptly (i.e., during SMTP negotiation) 413,105 spam e-mails with fixed addresses. The anti-spam filter CanIt-PRO classified 1,559,396 e-mails as spam, forwarding the remaining 429,928 legitimate e-mails to the recipients' mailboxes through the e-mail server Zimbra. Thus, ABL was responsible for rejecting a percentage of 20.94\% of the spam e-mails received by the university during this period. This percentage was probably improved when the recipients (i.e., e-mail users) started fine tuning their blacklists by means of the use of the blacklist interface (Section~\ref{sec:interface}). Figure~\ref{fig:deploy} plots the daily number of spam e-mails rejected promptly (i.e., during SMTP negotiation) by ABL.

\section{ABL versus anti-spam mechanisms}

The best current commercial and open-source anti-spams (ASes) make use of three main mechanisms --- address lists, SMTP extensions, and machine learning models --- for e-mail classification.

There are three types of address lists --- whitelists, greylists, and blacklists. Whitelists \cite{levi10} contain addresses or domains of legitimate e-mail senders or servers. Whitelists are not scalable, i.e., they will grow largely in size with the full adoption of the Internet Protocol version 6 (IPv6) \cite{deer17}, causing a serious decay in performance to access them.

Greylists \cite{kuch12}, \cite{harr03} 
store 3-tuples, built from the e-mails received by a server. Each 3-tuple\footnote{A 3-tuple is also referred to as {\it triplet}.} is composed of three types of information --- IP address of the host attempting the delivery, envelope sender address, and envelope recipient address. Every time a server receives an e-mail, it extracts its corresponding 3-tuple and checks if it is present on its greylist. If it is, the server classifies the e-mail as legitimate and dispatches it to the recipient's mailbox. Otherwise, it temporarily declines the e-mail and waits for its re-transmission. If it receives the re-transmission, the server classifies the e-mail as legitimate and dispatches it to the recipient's mailbox. Otherwise, it discards the received e-mail. The policy on which greylists are based assumes that legitimate e-mail servers maintain e-mail queues and have re-transmission policies in case of temporary errors. However, spammers can also implement this same re-transmission policy in order to confuse servers that use greylists.

Blacklists \cite{levi10} are lists which contain addresses or domains of suspicious e-mail senders or servers. Blacklists have four serious drawbacks. Firstly, they may not be updated as fast as the spammers change their sender addresses or domains. Secondly, a legitimate e-mail service provider runs always the risk of having any of its addresses (or domains) unduly inserted into one or more blacklists, and that causes it considerable annoyances \cite{harb20,croc22,huck19,gard20}. Thirdly, such as whitelists, blacklists are not scalable as well, i.e., they will grow largely in size with the full adoption of the IPv6, causing a serious decay in performance to access them. At last, the most trusteable blacklists are managed by operators who charge for their use. Thus, organizations have to pay annual fees for using ASes that use them.

The most employed extensions of the Simple Mail Transfer Protocol (SMTP) \cite{klen15} are the Sender Policy Framework (SPF) \cite{kitt14}, Domain-Keys Identified Mail (DKIM) Signatures \cite{croc11}, and Domain-based Message Authentication, Reporting, and Conformance (DMARC) \cite{kuch15}. SPF policy requires querying the records of the Domain Name System (DNS) servers of the domain to check whether or not the e-mail server that sent the e-mail has permission from the domain to send e-mails. SPF is incompatible with e-mail forwarders as well. DKIM, in turn, requires the use of encrypted signatures to validate the e-mail sender, and DMARC builds on SPF and DKIM. Thus, all three SMTP extensions generate additional expenses, both in terms of computation and time. In addition, they show vulnerabilities when implemented together in a same e-mail server. Indeed, Chen et al.\ \cite{chen20} discovered various vulnerabilities in e-mail servers of ten popular e-mail providers. All these state-of-the-art servers implement SPF, DKIM, and DMARC.

Machine learning models are artificial intelligence models capable of being continuously trained on pattern databases. Thus, they require periodic training on patterns composed of e-mail features --- sender, recipient, header, subject, body, attachments, among others --- in order to be able to accurately classify new e-mail patterns.

In addition to the deficiencies mentioned, the three main mechanisms are also passive. They require that spam e-mails be received entirely before evaluating and discarding them, wasting the computational and network resources of the user or of the institution. Moreover, they do not penalize the spammer, returning to him/her his/her spam e-mails, and thus the spammer goes on keeping the recipient in his/her distribution lists.

The Active Blacklist (ABL), in turn, assumes that each recipient (i.e., e-mail user) knows, with absolute certainty, how to classify the e-mails he/she receives as ham or spam. He/she is able to correctly assemble his/her blacklist with the sender addresses used by spammers. Thus, ABL does not need to receive and store the spam e-mail addressed to the recipient if its sender address is on the recipient's blacklist. Therefore, ABL can promptly reject the spam e-mail during SMTP negotiation.

ABL is based on a modification of the Simple Mail Transfer Protocol (SMTP). It differs from usual passive blacklists in that it produces three advantageous consequences. Firstly, it promptly rejects, during SMTP negotiation, the spam e-mails thus defined by each e-mail user of an organization, avoiding a waste of the computational and network resources of the organization. Secondly, it returns the spam e-mails to the spammer, penalizing him/her, for his/her server will use more computational and network resources to handle the rejected spam e-mails. Thirdly, owing to the cost of the refusal, the spammer usually removes the user e-mail address from his/her distribution lists.

ABL targets specifically spam messages with fixed sender addresses. Such spam messages represent a significant amount of the spam that arrives at the Federal University of Itajub\'{a}. During the first sixty-one days of the operational period of ABL at the university, it was responsible for rejecting a percentage of 20.94\% of the spam e-mails received by the university during these days.

\section{Conclusion}

This paper presents the Active Blacklist (ABL). It is based on a modification of the SMTP. It was implemented in the Mail Transfer Agent (MTA) Postfix of the e-mail server Zimbra in order to be assessed experimentally. ABL allows recipients to mark which e-mail sender addresses are undesired.

ABL returns promptly (i.e., during SMTP negotiation) spam e-mails to senders, as if the recipient did not exist. Hence, not only the computational resources of the destination server are preserved, but also those of the spammer's server are burdened.

A number of experiments were carried out to assess the computational costs of the modified server Zimbra in relation to those of the original server Zimbra. When the host sends ham e-mails, the additional cost to the modified server is insignificant. When the host sends spam e-mails, the modified server reduces significantly its CPU usage as well as the maximum number of e-mails in its queue {\it incoming}. Transmission (Tx) of bytes through its network interface increases owing to the immediate return of the spam e-mails to the host. Nonetheless, this increase is offset by the reduction in the number of bytes received (Rx) in its network interface.

It was also verified through the experiments that there is a considerable computational cost on the host when it sends spam e-mails. When sending spam e-mails only, its average rates of CPU usage, of RAM memory usage and of bytes received (Rx) through its network interface increase in 93.48\%, 98.59\% and in almost 40,000\% respectively, in relation to the average rates obtained when only ham e-mails are sent. Nonetheless, when sending spam e-mails only, the average rate of bytes transmitted (Tx) through its network interface decreases in 50\% in relation to the average rate obtained when only ham e-mails are sent.

ABL was deployed in the production environment of the Federal University of Itajub\'{a}, Brazil, and was operational from July-2015 to July-2019. It is not presently in use, for the university e-mail service was migrated to the G-Suite platform of Google in August-2019. During the first sixty-one days of its operational period, ABL was evaluated. The evaluation revealed that ABL was responsible for rejecting a percentage of 20.94\% of the spam e-mails received by the university during these days.

There are some possibilities for future work. First, testing other MySQL engines in order to investigate whether they perform better than the InnoDB. Second, implementing a service to remove expired e-mail addresses from the {\it BlackList\/} table. The expiry term may be defined by the network administrator. Third, letting the search in the {\it BlackList\/} table be also made by any string of the sender's address (e.g., the sender's domain).


\bibliographystyle{plain}
\bibliography{bib}

\end{document}